# High density two-component glasses of organic semiconductors prepared by physical vapor deposition.

Yejung Lee[1], Shinian Cheng[1], and M. D. Ediger[*,1]

[1]*Department of Chemistry, University of Wisconsin-Madison, Madison, Wisconsin, 53706, USA*

## Abstract

Physical Vapor Deposition (PVD) is widely utilized for the production of organic semiconductor devices due to its ability to form thin layers with exceptional properties. Although the layers in the device usually consist of two or more components, there is limited understanding about the fundamental characteristics of such multi-component vapor-deposited glasses. Here, spectroscopic ellipsometry was employed to characterize the densities, thermal stabilities and optical properties of co-vapor deposited NPD and TPD glasses across the entire range of composition. We find that co-deposited NPD and TPD form high density glasses with enhanced thermal stability. The dependences of density and stability upon substrate temperature are correlated, and the birefringence of the co-deposited glasses is determined by the reduced substrate temperature of mixtures. Additionally, we observe that the transformation of a highly stable and dense two-component glass into its supercooled liquid initiates from the free surface and propagates into the bulk at constant velocity, like single component PVD glasses. All these features are consistent with the surface equilibration mechanism.

## TOC

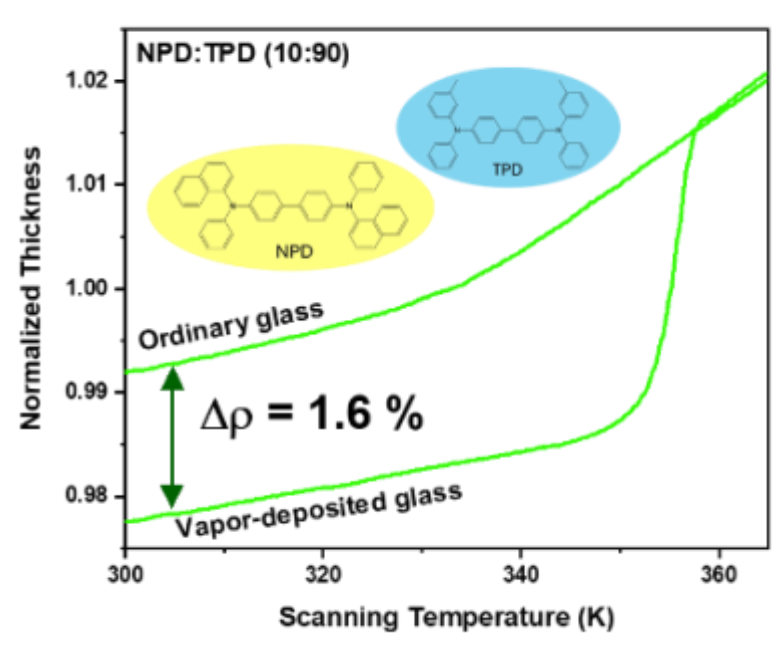

The past decade has seen great advances and rapid growth of organic semiconductors. Especially, organic light-emitting diodes (OLED) have been successfully commercialized as displays in cellular telephones and televisions. As a result, a fundamental understanding about the physical state of OLED layers is of great interest.(*1-5*) Emissive and conductive layers in these devices are composed of glassy solids even though a crystalline film may have higher charge mobility. One key advantage of glassy OLED layers is macroscopic structural uniformity; in contrast to poly-crystals, flat surfaces and the absence of grain boundaries enable consistent performance over a wide area.(*6*) Moreover, glassy solids readily accommodate a second component allowing the good dispersion of an emitter in the host matrix, which is necessary to avoid concentration quenching. Glasses are inherently in non-equilibrium states and thus different properties of glasses can result depending upon the manner in which the glass is prepared. (*6-8*)Conventionally, the glassy layers in OLEDs are formulated either by solution processing or by vapor deposition.

Physical vapor deposition (PVD) can prepare glassy films with a range of characteristics depending upon deposition conditions.(*1, 6, 9*) In particular, PVD can produce ultra-stable and also high density glasses.(*6, 8, 10-13*) Previous studies on vapor-deposited glasses have shown high kinetic stability(*8, 11, 12, 14-18*), low enthalpy(*8, 15, 19, 20*), high mechanical moduli(*21-23*), high chemical stability (*24*) and high photo-stability(*25*). These properties can be explained by the surface equilibration mechanism, which is based on high molecular mobility at the free surface relative to the bulk. (*6, 8, 26, 27*)Consistent with this mechanism, the density of a vapor-deposited low molecular weight system, ethylbenzene, was shown to approach the density of the ideal glass.(*28*) Fakhraai and coworkers have recently reported that a vapor-deposited glass of TPD can have significantly high density when it is ultrathin (thinner than 50 nm) (*10*)or deposited on a soft substrate(*29*). Furthermore, the structure of molecules in the as-deposited glasses can be anisotropic whereas liquid-cooled glasses are usually isotropic.(*1, 2, 9, 12, 30, 31*) Dalal et al. demonstrated that three linear-shaped organic semiconductors (NPD, TPD and DSA-Ph) have tunable molecular orientation depending on the ratio of the substrate temperature ($T_{sub}$) to the glass transition temperature ($T_g$) during deposition.(*12*) The orientation of disk-like molecules can also be systematically controlled with the substrate temperature.(*14*)

These exquisite properties of vapor deposited glasses are important in electroluminescent (EL) devices, such as OLEDs, which are commonly prepared by PVD.(*1-5, 32-36*) The stability and density of the as-deposited glassy films contributes to the device durability and performance.(*37*) Rafols-Ribes et al reported that OLEDs prepared by vapor deposition at 0.85 $T_g$ (which produces the highest density glasses) have up to 5 times longer lifetime in comparison to deposition at other temperatures.(*5*) Also, a recent study from Esaki et al demonstrates that a 1 % increase in density is correlated with a 25-fold increase of charge carrier mobility for PVD glasses of NPD.(*33*) From the perspective of EL technologies, the tunability of molecular orientation by vapor deposition has an important impact on light outcoupling.(*2, 9*) The transition dipole orientation of the emitting molecule is one of the most important factors influencing the performance of devices.(*38*) Horizontally oriented emitter molecules facilitate the light traveling out of the device and maximizes the outcoupling efficiency in OLEDs relative to dipoles in a vertical or isotropic orientation.(*3, 4, 9*)

Most previous studies about the stability and structure of PVD glasses have focused on single component systems and much less is understood about multi-component glasses. Since some layers in OLEDs are mixture of two or more molecules, it is important to elucidate whether multi-component PVD glasses also have those desirable properties found for single component glasses. When two or more molecules are co-deposited, the different $T_g$ values, or equivalently, different degree of mobilities, and interaction between the molecules can influence equilibration during the film growth. A recent study by Cheng et al. found that co-deposition of organic semiconductors, TPD and m-MTDATA, in 50:50 ratio yielded glasses with exceptional kinetic and thermodynamic stability; these results were interpreted based on the surface equilibration mechanism.(*39*) In a second paper, Cheng et al. identified an additional five co-deposited systems of organic semiconductors (all 50:50 mixtures) that exhibited high kinetic stability in spite of differences in $T_g$ of up to 96 K.(*40*) In contrast, Ki et al reported that co-deposited glasses of Liq and Bphen did not exhibit high kinetic stability.(*37*) These results point out an incomplete understanding of co-deposited PVD glasses, and some important features of vapor-deposited glasses, such as density relative to the liquid-cooled glass, have not ever been reported to our knowledge. Furthermore, previous work on the co-deposited PVD glasses has focused on equimolar mixtures but other regimes are also of interest, such as low concentration dopants in a host matrix, as these are typical for emitter layers in OLEDs.

In this paper, we investigate co-deposited organic semiconductor glasses of NPD and TPD. These linear molecules have similar size and a 35 K difference in $T_g$ values. For context, we performed differential scanning calorimetry (DSC) on bulk mixtures of this binary system; a single $T_g$ is observed for every composition and the phase diagram indicates that this mixture is thermodynamically ideal. We utilize spectroscopic ellipsometry to determine the density, thermal stability, and birefringence of co-deposited glasses across the entire range of composition, and we vary the substrate temperature across a wide range at each composition. We find that the onset temperature, $T_{onset}$, for transformation to the liquid is elevated by up to 7 % compared to the liquid-cooled glass, and the density of the as-deposited glasses is up to 1.6 % higher. To our knowledge, this is the first measurement of relative density for a co-deposited PVD glass. For all compositions, the dependence of the density on $T_{sub}/T_g$ is very similar to that of single component glasses, showing that the surface equilibration mechanism describes important features of the co-deposited NPD:TPD system. The birefringence depends upon $T_{sub}/T_g$ for each composition in a manner similar to that for PVD glasses of the two pure components. We also performed isothermal transformation experiments on a highly stable, co-deposited glass into the liquid. We observed that the transformation initiates from the free surface and propagates into the stable glass at a constant velocity.

Figure 1 shows spectroscopic ellipsometry data portraying the thickness changes upon heating and cooling for NPD:TPD (10:90) glasses vapor-deposited at different temperatures from 230 K to 330 K. A rate of 5 K/min was utilized for both heating and cooling and three temperature cycles were performed in order to characterize the thermal stability and the relative density of as-deposited glass. Upon heating, the thickness of the PVD glasses slightly increases linearly due to thermal expansion, indicated by the slope, while maintaining their molecular packing as vapor-deposited glasses. At the onset temperature ($T_{on}$), a significant jump-up in thickness is observed as the glass begins the transformation to the supercooled liquid. After the transformation to the liquid state is complete, the thickness data joins to the supercooled liquid line and increases linearly with a larger thermal expansion coefficient. Subsequently, as the temperature lowered, the thickness of the supercooled liquid decreases gradually, retracing the thickness data obtained during heating. Upon further cooling, the transition to the liquid-cooled glass occurs as noted by the deviation from the extrapolated behavior of the equilibrium supercooled liquid (denoted as $T_g$). In the 2nd and 3rd temperature cycles (data not shown), the thickness changes overlap those observed for the first cooling cycle except for the small, expected hysteresis near $T_g$.

To compare these data sets, the thickness data was normalized at the measured $T_g$ value. The glass transition temperatures for the liquid-cooled glasses of all the 6 samples shown in Fig 1 are expected to be the same since the thermal history of the sample is erased in the first heating ramp. We observed that all the $T_g$ values are equal to 334 K with a standard deviation of less than 1 K. We interpret this small variance to indicate that all six deposited glasses have the same composition, to within a mass ratio of 0.02.

One key aspect of Figure 1 is that as-deposited glasses of NPD and TPD are generally thinner than the corresponding liquid-cooled glasses, and thus have higher densities. It is convenient to use the fictive temperature, $T_f$, to quantify glass properties such as density. Qualitatively, $T_f$ represents the temperature where a liquid-cooled glass would leave equilibrium, in order to achieve the same density of as-deposited glass. Quantitatively, $T_f$ is determined by where the extrapolated supercooled liquid line meets with the as-deposited glass line, as indicated in Figure 1.

A second key aspect of Figure 1 is that the onset temperature, $T_{on}$, is systemically larger than $T_g$ for the liquid-cooled glass, by up to 19K. These indicates the co-deposited glasses of NPD and TPD have substantially higher kinetic stability than the liquid-cooled glass, because greater thermal energy is required to dislodge the molecules from their glassy packing.

Figure 2 displays the results of $T_{onset}$, $T_f$ and relative density as a function of substrate temperature, covering the entire composition range from pure TPD to pure NPD. Results for NPD:TPD composition other than 10:90 were extracted using the same approach shown in Figure 1; all the co-deposited glass data shown in solid symbols. Single component glasses of neat NPD and TPD deposited at various substrate temperatures are also plotted in Figure 2 (open symbols).

In all cases, the data has been normalized to the $T_g$ value for liquid-cooled glass of the same composition.

Figure 2 shows that, for every composition of co-deposited NPD:TPD glasses, the most stable and highest density glasses are formed where the deposition temperature is between 0.8 to 0.9 of $T_g$ for the liquid-cooled glass. The co-deposited glasses are up to 1.5 % more dense than the liquid-cooled glasses, indicating efficient molecular packing. A significant correlation between higher thermal stability ($T_{onset}/T_g$) and denser molecular packing is also observed in every composition.

The results shown in Figure 2 are quite similar to those previously reported single-component vapor-deposited glasses of organic semiconductors(*11, 12, 15*), and can be interpreted using the surface equilibration mechanism that has proven useful for single-component systems.(*11, 12*) During deposition, the molecules at the surface have much higher mobility than molecules in the bulk of the glass.(*26, 27, 41*) This surface mobility allows freshly-deposited molecules to equilibrate towards low- energy packing configurations. Subsequent deposition traps this packing into the bulk glass, explaining how vapor-deposited glasses have greater density and greater kinetic stability in comparison to liquid-cooled glasses. The substrate temperature controls the stability of the vapor-deposited glass through a competition of thermodynamics and kinetics. Lower substrate temperatures can produce glasses with high stability if the deposition rate is low enough to allow substantial equilibration. At a fixed deposition rate (as used in these experiments), equilibration is nearly complete for the highest substrate temperatures but is much less effective at the lowest substrate temperature; this accounts for the presence of an optimal temperature range, near 0.85 $T_g$, for preparing high density and high thermal stability glasses. A test of the surface equilibration mechanism is shown in the bottom panel of Figure 2. At substrate temperatures above about 0.92 $T_g$, for all the co-deposited glasses, we observe that the data is consistent with the $T_f = T_{sub}$ line shown in the figure. This data indicates that the equilibrium density has been attained in these depositions, as expected when surface mobility is very high.

Figure S2 shows that the refractive index (n) at 650 nm, as a function of temperature of liquid-cooled glasses of NPD and TPD, as obtained from in-situ ellipsometry. Ellipsometry is a powerful method for determining the optical constants of organic semiconductors and these are important in device design. For each composition, the refractive index values shown represent an average of results for 5 samples. As temperature increases, the refractive index values decrease, which is consistent with the increase of thickness as shown in Figure 1. The arrows in the figure show the glass transition for the NPD and TPD glasses prepared in different compositions, as identified by the change in slope. These results provide a check on the Tg values determined from the sample thickness (as shown in Figure 1). For all compositions, the difference of the Tg values obtained from the two methods was less than 0.7 K.

Birefringence values ($\lambda$ = 650 nm) for as-deposited NPD:TPD glasses are presented in Figure 3(a) as a function of the normalized substrate temperature, for all compositions including the pure components. Birefringence is determined by the difference of extraordinary (out-of-plane) and ordinary (in-plane) indices of refraction ($n_z$-$n_{xy}$) and serves as a proxy for average molecular orientation. As an example, Dalal et al. showed for NPD that the birefringence and molecular orientation associated with the transition dipole moment follow very similar trends as a function of substrate temperature.(*12*) For NPD and TPD, negative birefringence values indicate that the molecular long axis tends to lie in the plane of the sample (consistent with face-on packing) while the (small) positive birefringence values indicate a slight tendency for the molecular long axis to lie along the surface normal (end-on packing). Figure 3(a) demonstrates that, when plotted as a function of $T_{sub}/T_g$, vapor-deposited glasses of NPD, TPD, and their mixtures have very similar birefringence values. As NPD and TPD have similar molecular structures, we interpret this result to mean (1) that the average molecular orientation depends upon substrate temperature in the same way for the pure materials and the mixed glasses, and (2) that NPD and TPD have similar molecular orientations in the co-deposited glasses.

A recent paper from Cheng et al. developed a prediction for birefringence of co-deposited glasses (*40*) based upon the surface equilibration model where Δn represents the birefringence and φ means volume fraction of each components:

$$\Delta n_{AB}\left(\frac{T_{sub,AB}}{T_{g,AB}}\right) = \varphi_A * \Delta n_A\left(\frac{T_{sub,A}}{T_{g,A}}\right) + \varphi_B * \Delta n_A(\frac{T_{sub,B}}{T_{g,B}})$$

This approach assumes that the orientation of each component in the mixture is controlled by the normalized substrate temperature but is otherwise independent of composition. Cheng et al. compared this prediction to the observed birefringence of 50:50 co-deposited mixtures of organic semiconductors and found good agreement. In Figure 3(b), we test the proposed equation for NPD:TPD glasses over a wide range of compositions. In all cases, the experimental data (solid points) are in quite good agreement with the prediction (dashed lines). In our comparison, we utilized the weight fraction of the two materials rather than the volume fraction but for the NPD/TPD system this is inconsequential. It is notable that the model is working not only for the 50:50 mixtures (as expected based upon the results of ref. 40) but also for the dilute mixtures.

Figure 4 characterizes the interaction between NPD and TPD as determined by differential scanning calorimetric (DSC) measurements, and we find results consistent with an ideal mixture. Figure 4 (a) and (b) respectively depict the melting temperatures and glass transition temperatures of mixtures of NPD and TPD. Physically mixed NPD and TPD crystals were heated to measure melting temperatures, and these values were fitted with Schroder-Van Laar equation (dashed lines in panel a). Here, $T_l$ denotes the liquidus temperature, where the material is completely liquid above this temperature and $T_e$ means eutectic temperature. As this equation assumes ideal mixing behavior, the good fit is consistent with an ideal mixture at temperatures near the melting points, indicating that bulk binary blends of NPD and TPD are expected to be miscible at every composition. In addition, subsequent temperature cycles in the DSC experiments showed a single glass transition, with $T_g$ values as plotted in Figure 4 (b). The dashed line is Gordon-Taylor (GT) model(*42-44*), assuming that $\Delta V_{mix} = 0$. The predicted $T_g$ curves in terms of the model reasonably match the experimental measurements. This result is consistent with the view that NPD and TPD form an ideal solution near $T_g$. Figure 4(c) strengthens our conclusions by calculating the activity coefficient for TPD as a function of composition, using the data from Figure 4(a). The calculated activity coefficients are all close to unity, consistent with an ideal solution. $\Delta H_{m,i}$, $x_i$ and $\gamma_i$ denote the enthalpy of fusion, molecular fraction and activity coefficient of component i, respectively, and R is the ideal gas constant.

$$-\ln x_i \gamma_i = \Delta H_{m,i} \left( \frac{1}{T_l} - \frac{1}{T_{m,i}} \right) \frac{1}{R}$$

Single-component highly stable glasses transform into the supercool liquid upon heating by a route not observed for other glasses(*20, 45*), and we tested whether this is also true for highly stable two-component glasses. When single-component stable glasses are annealed above $T_g$, the transformation initiates from the free surface because molecular mobility is largest there, and a front is observed because “nucleation” of the supercooled liquid in the bulk of the glass is very slow.(*45-47*) In order to test the transformation mechanism of a two component PVD glass, we deposited at 50:50 NPD:TPD glass at 315 K (0.91 $T_{g,mix}$), and then subsequently isothermally annealed at 360 K ($T_{g,mix}$ +14 K). It is worth noting that the as-deposited glass is 1.23 % denser than the corresponding liquid-cooled glass, and the annealing temperature was selected to be above $T_{g,mix}$ but below $T_{onset}$. Figure 5 displays ellipsometry data showing the time evolution of as-deposited glass during isothermal annealing. When fitted to a three-layer model, the position of transformation fronts could be identified near the free surface and near the substrate.(*46*) The series of many independent data sets shows that these fronts move towards the center of the film at a constant rate. Along with that, the corresponding MSE data for this three-layer model and homogeneous model are shown together in Figure 5. When comparing the MSE, the homogenous model based on single layer have MSE over 15 whereas two growth front model fitted well with less than 4 MSE values, which confirms the as-deposited glass transforms by a front growth mechanism. Based on the previous findings, this is consistent with high density and ideal mixing behavior of NPD and TPD glasses. Thus, we show that transformation by fronts also occurs in multi-component PVD glasses.

Through the use of in-situ spectroscopic ellipsometry, we have presented here the first evidence that co-deposited glasses can have densities up to 1.6 % higher than the corresponding liquid-cooled glass mixture. In addition, we have shown that as-deposited thin glassy films of NPD and TPD have substantially enhanced thermal stability. The high density and enhanced stability of the co-deposited glasses studied here equals or exceeds that reported for single-component PVD glasses.(*11, 12, 14, 21, 24, 25*)

In previous work on single-component systems, high density has been correlated with important improvements in material properties, and we anticipate that improved properties will also be found for dense co-deposited glasses. Esaki et al previously reported that 1 % denser glasses of neat NPD possess 25 times higher charge carrier mobility.(*33*) Given this result, the 1.5 % density increase reported here may have a very substantial impact on charge mobility. Qiu and co-workers found that the vapor deposited Disperse Orange 37 (DO37) was much more photostable than a liquid-cooled glass of the same material, and the photostability was strongly correlated with the density. The 1.03 % density increased vapor-deposited DO37 glass was 50 times more photostable than the liquid-cooled glass.(*25*) Additionally, Qui et al. reported that dense glass packing of as-deposited indomethacin (a carboxylic acid) slows the chemical reaction with ammonia gas.(*24*) Again, a good correlation between higher glass density and reduced reaction rate was observed in this study. Several previous studies have shown that single-component, high density PVD glasses have enhanced mechanical properties, including an increase in modulus and hardness.(*21-23*) We look forward to future investigations of the material properties of co-deposited organic semiconductors glass mixtures such as NPD:TPD.

In previous reports on multi-component PVD glasses, ellipsometric data has only been utilized to characterize kinetic stability; no previous reports provide density measurements for multi-component PVD glasses. In contrast, for single component PVD glasses, in-situ ellipsometry has frequently been employed for detecting the kinetic stability and density increases for as-deposited thin glassy films, in comparison to liquid cooled glasses.(*1, 10-12, 14, 24, 25, 28*) In our experience using ellipsometry to study other pairs of organic semiconductors(*39, 40*), we have observed that the onset temperature was reproducible, whilst the relative density was not. We assume that this somehow results from the failure of the ellipsometric model for multi component films. This is surprising, since we expect an effect medium approximation to be suitable for any well-mixed organic thin films.

We chose to study NPD:TPD mixtures with the idea that two features might lead to reliable density measurements: 1) these two molecules have very similar molecular size and shape, and 2) PVD glasses of these two molecules show very similar birefringence values as a function of substrate temperature(*12*), indicating similar packing arrangements in the glassy state. The birefringence results for NPD:TPD mixtures shown in Figure 4 confirm this second point. Out investigations also reveal that NPD and TPD mixtures are thermodynamically ideal (or very close to it), as shown in Figure 4, and this may also be an important feature explaining why highly reliable ellipsometry data could be obtained for this pair. We do not have a definitive explanation for the success of ellipsometry measurements for this particular mixture. We hope that future work can successfully obtain density measurements on other co-deposited glasses, so that the generality of the results reported here can be investigated.

For co-deposited NPD:TPD, we also investigated the transformation mechanism that allows the supercooled liquid to be formed from a highly stable glass upon heating. Similar to highly stable single-component PVD glassy thin films, during isothermal annealing we observed for NPD:TPD a constant velocity propagating front that initiates at the free surface. This is an important observation for organic semiconductors because thin glassy films are often stacked in layers, such that no free surfaces are present. For single-component PVD glasses, it has been reported that eliminating the free surface for a thin film can dramatically increase kinetic stability.(*48*) This means that, in a device geometry, co-deposited NPD:TPD would be even more stable than what we report in Figure 2. We anticipate that other co-deposited semiconductor mixtures will transform via propagating fronts and thus share this extra source of stability.

Beyond information about the density and stability, our ellipsometry studies of NPD:TPD also reveal important information about molecular orientation in co-deposited PVD glasses. Previous work has shown that PVD glasses of single-component systems are often anisotropic, and the prevailing molecular orientation can be a strong function of substrate temperature.(*11, 12, 14*) The anisotropy of PVD glasses has been explained in terms of the surface equilibration mechanism. Generally, low temperature depositions result in face-on packing. Higher temperature depositions allow the opportunity for equilibration at the free surface, towards whatever structure is favored thermodynamically. Several recent works have used the surface equilibration mechanism to explain dipole orientation in single-component PVD glasses.(*30, 35*)

In this work, we show that molecular orientation (as inferred from the birefringence) in NPD:TPD mixtures with different compositions, is very similar to orientation in the neat materials, as long as the comparison is made at the same value of $T_{sub}/T_g$. Previous reports on PVD glasses of two or more components fall mostly into two categories: (1) focusing only on dilute mixtures (less than 10%) for emitter and host systems(*3, 4, 31, 49, 50*), or (2) PVD glasses with 50:50 ratio(*39, 40*). In terms of PVD glasses with 50:50 mole ratios, our efforts extend the recent work of Cheng and coworkers(*39, 40*), by studying dilute mixtures as well as 50:50 mixtures. We showed that the birefringence mixing rule that they proposed (based upon the surface equilibration mechanism) works in both regimes. In terms of PVD glasses with one dilute component, there has been extensive work using fluorescence methods to measure emitter orientation of transition dipole moment.(*3, 9, 31, 49, 50*) Much of that work has reported that $T_{sub}/T_g$ plays a critical role in determining molecular orientation(*11, 12, 32*), in a manner that is at least qualitatively expected based upon the surface equilibration mechanism. The results presented here provide the first test validation that the surface equilibration mechanism provides useful information about molecular orientation in co-deposited systems across the entire range of possible compositions.

We have shown that the NPD:TPD system can be described as thermodynamically ideal, and it is important to understand how the properties of co-deposited systems might be different if this is not the case. Cheng et al. recently studied co-deposited glasses of two organic semiconductors (TPD and mMT-DATA) with 50:50 composition.(*39*) They reported the formation of glasses with high kinetic stability, in common with results presented here which included a wider composition range. TPD and mMT-DATA were also shown to form a thermodynamically ideal mixture, so these results, together with the results presented here, are likely the generic behavior for ideal mixtures. Bishop et al. recently studied co-deposited glasses of DO37 and TPD using RSoXS and AFM. In contrast to the systems discussed above, DO37/TPD glasses show evidence for component segregation during deposition resulting in domains that can exceed 100 nm. DSC measurements for liquid-cooled glasses of that system show two $T_g$s, consistent with phase separation in the liquid state.(*51*) Additional experiments will be required to understand kinetic stability and molecular orientation in co-deposition glasses with component segregation. We note that our laboratory attempted to use ellipsometry to investigate TPD/mMT-DATA mixture and also DO37/TPD mixtures; our efforts did not yield reliable information about the relative densities of the as-deposited glasses.

In summary, this is the first report that physical vapor deposition can generate high density, mixed glasses of organic semiconductors, NPD and TPD. Co-deposited glasses of NPD and TPD across the whole range of composition have elevated thermal stability and high density; these quantities have a non-monotonic dependence on relative substrate temperature. In the entire range of compositions, including non-dilute and dilute mixtures, the trends were the same as for single component glasses, which is explained by the surface equilibration mechanism. Moreover, the as-co-deposited glasses are anisotropic and we are able to quantitatively predict the birefringence over the entire composition range, based upon the measured birefringence of the pure components. This work is a significant step towards understanding PVD glass mixtures and provides valuable insights for designing high performance electronic devices.

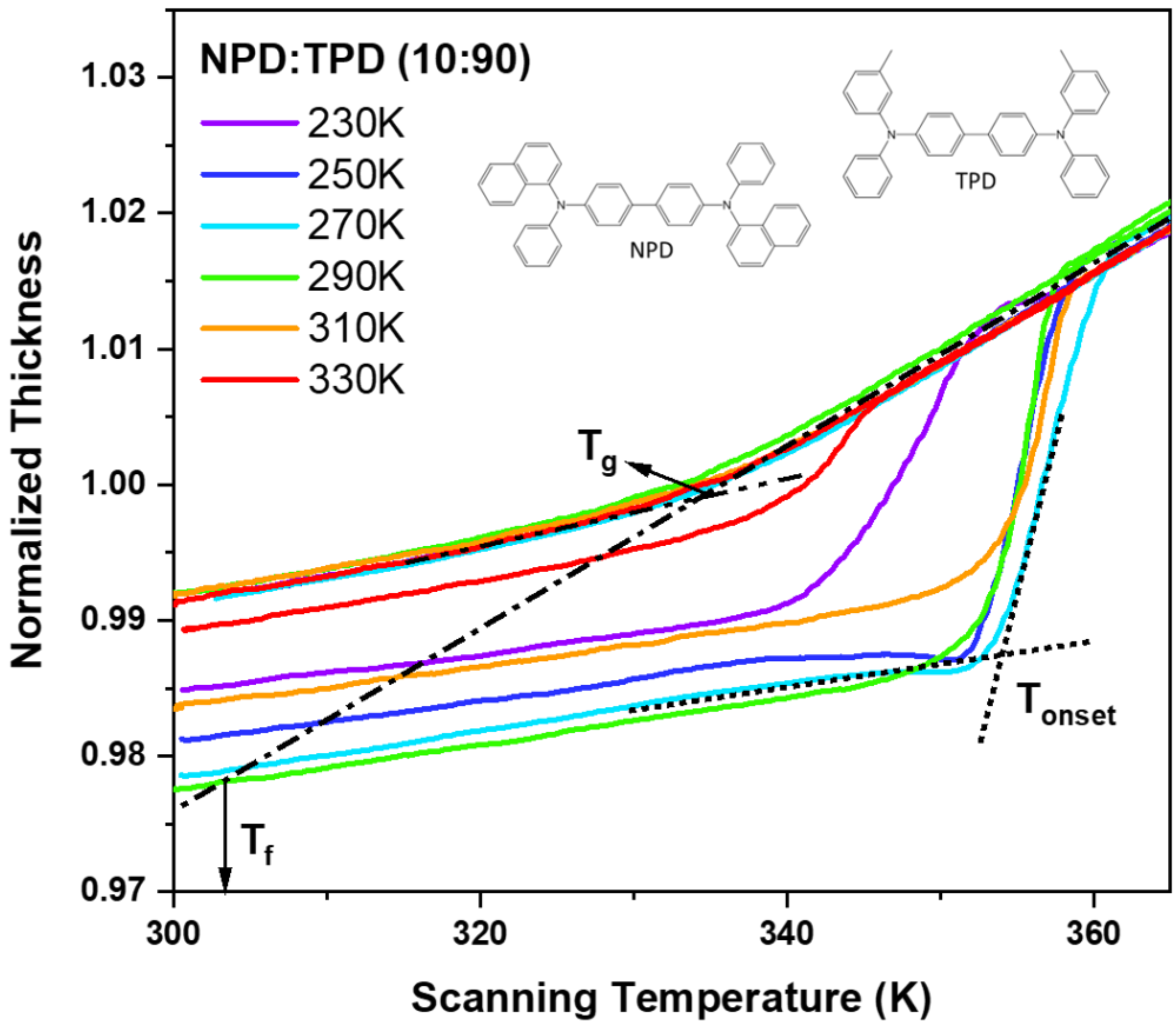

***Figure 1*** *Ellipsometric measurements of normalized density for as-deposited NPD:TPD (10:90) glasses and the corresponding liquid-cooled glasses, during heating and cooling at 5 K/min. This data was used to determine the onset temperature and the fictive temperature of as-deposited glasses and liquid-cooled glass transition temperature as well as their relative density. The substrate temperature during deposition was 230 K to 330 K as indicated by color. Thickness data is normalized to the thickness of the liquid-cooled glass at* $T_g$*; all samples were approximately 200nm thick. The chemical structures of TPD and NPD are shown as insets.*

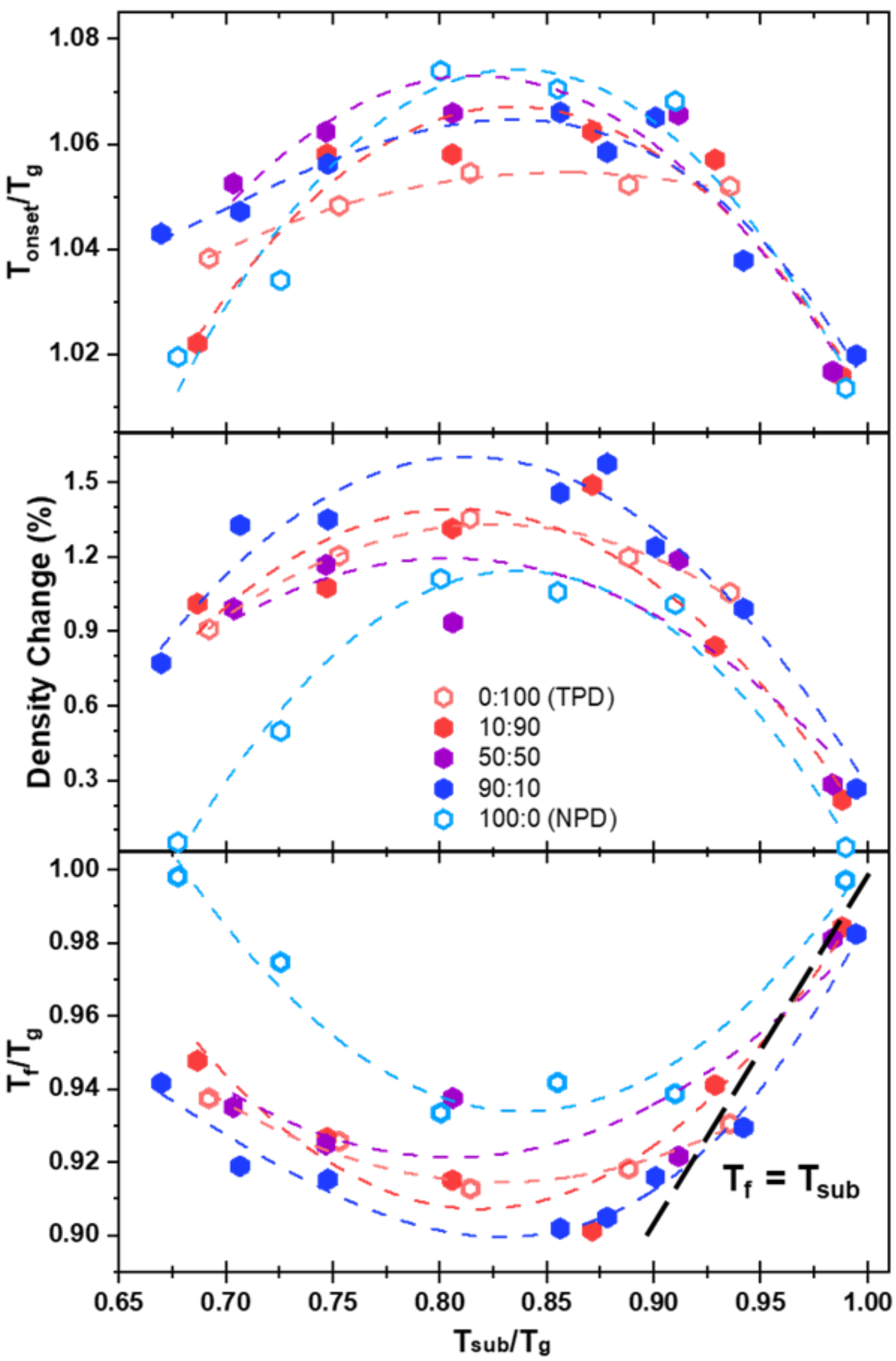

***Figure 2*** *Ellipsometric result of $T_{onset}/T_g$, Density change in % and $T_f/T_g$ for as-codeposisted NPD:TPD mixtures over the substrate temperatures. Open symbol indicates neat single component glasses of TPD (0:100) and NPD (100:0) for reference while the closed symbols are for co-vapor deposited NPD:TPD in different compositions as represented in the legend. The $T_g$ used here is all from corresponding data.*

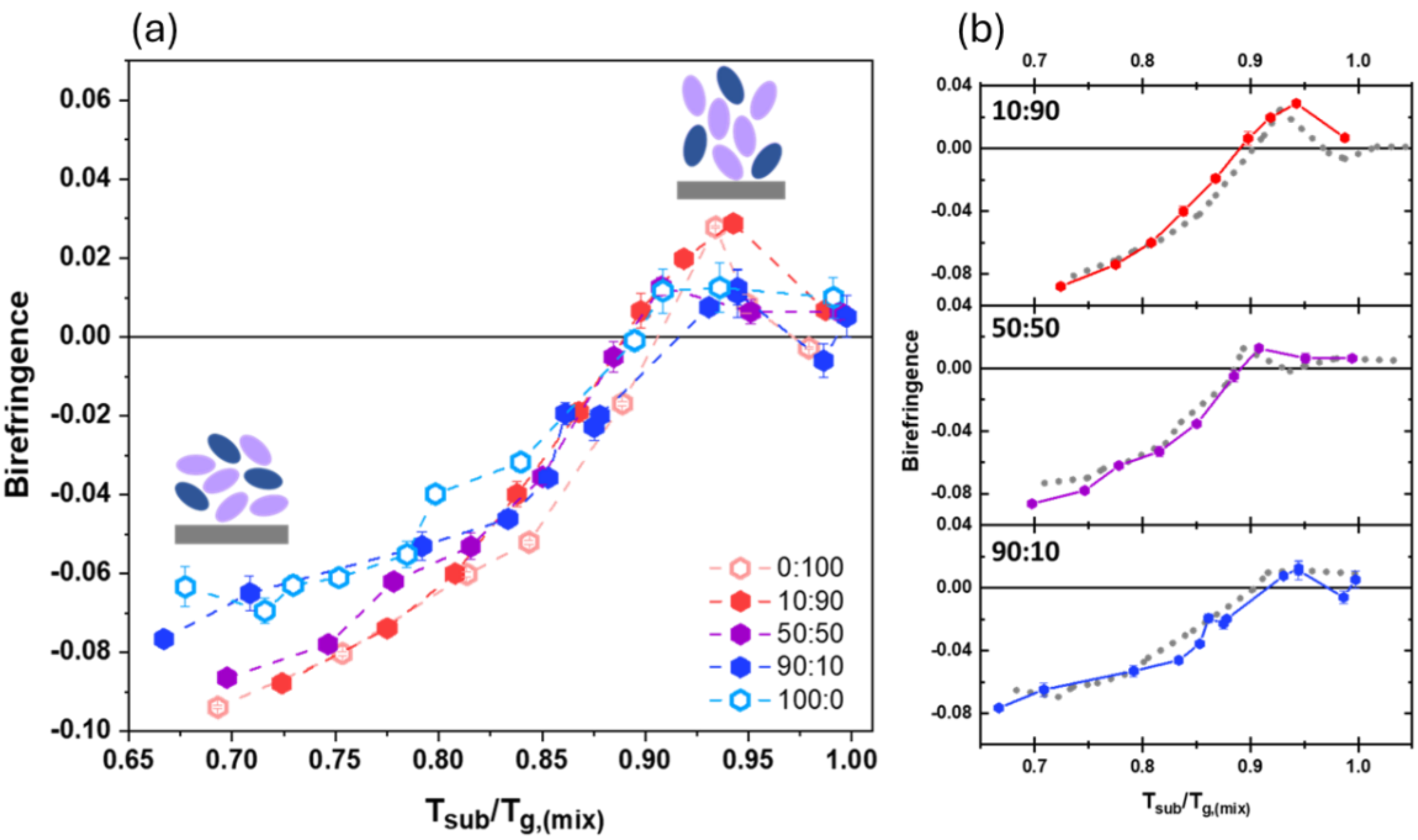

***Figure 3*** *Optical birefringence at 650 nm of vapor-deposited NPD:TPD glasses in entire ratio by varying the substrate temperature, which was measured at room temperature by ellipsometer. The right panel compares the model suggested in the previous paper, Cheng et al (2024) and experimental measured results. The grey colored line is from model, whereas the colored lines are same as in the left panel.*

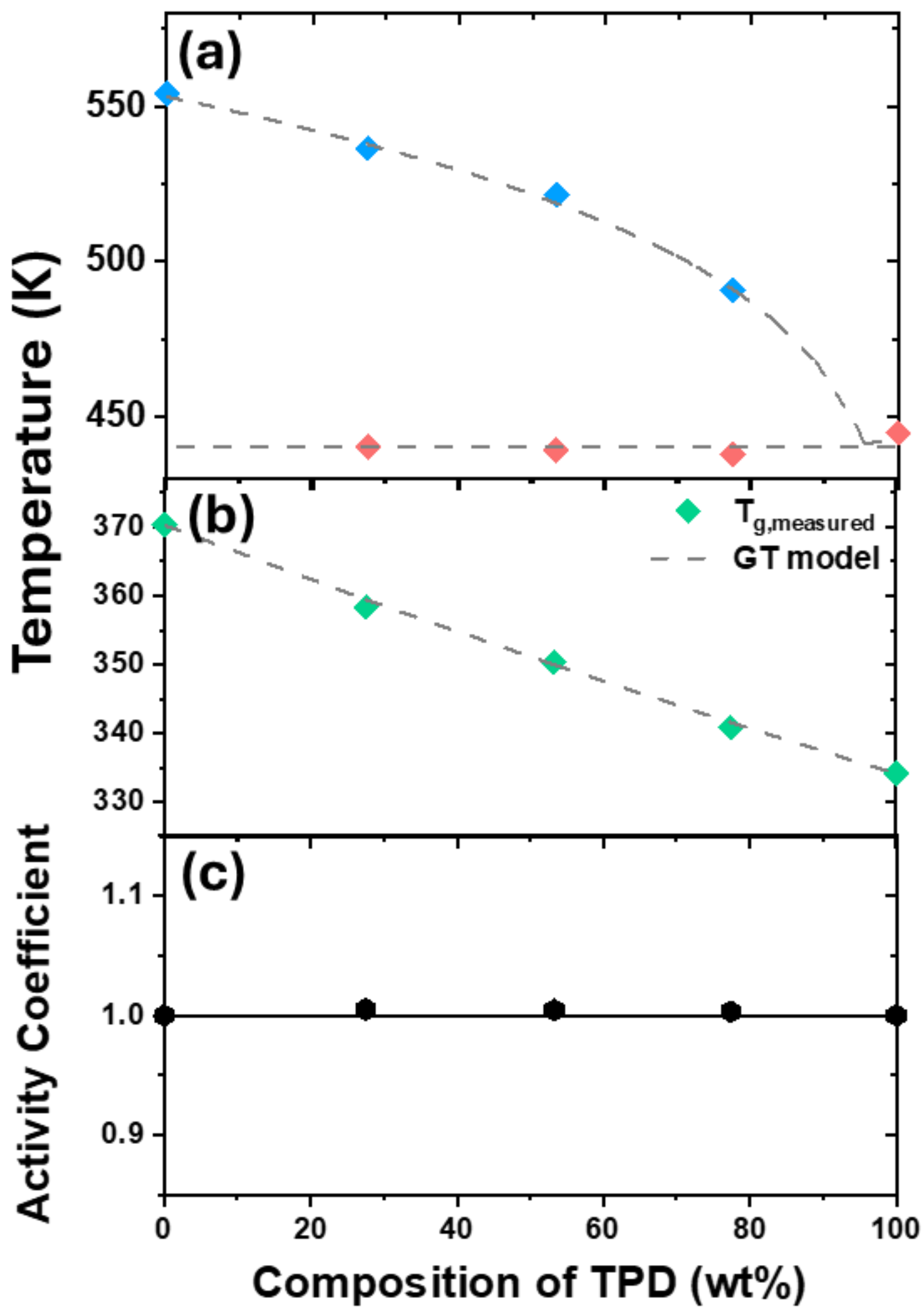

***Figure 4*** *DSC results of bulk mixture of TPD and NPD. (Top) Melting temperature and eutectic temperatures ($T_e$) at different composition in solid symbols. $T_l$ denotes the temperature where the whole material melts to liquid. The grey dashed line is Schroder-Van Laar equation, which is based on the ideal mixing liquid. (Middle) Glass transition temperatures of bulk mixtures of NPD and TPD. The predicted value of Gordon-Taylor model is in grey line was also plotted together to show that the mixture of NPD and TPD is ideally mixed. (Bottom) The calculated activity coefficient (equation shown in the text) as a function of TPD mass concentration.*

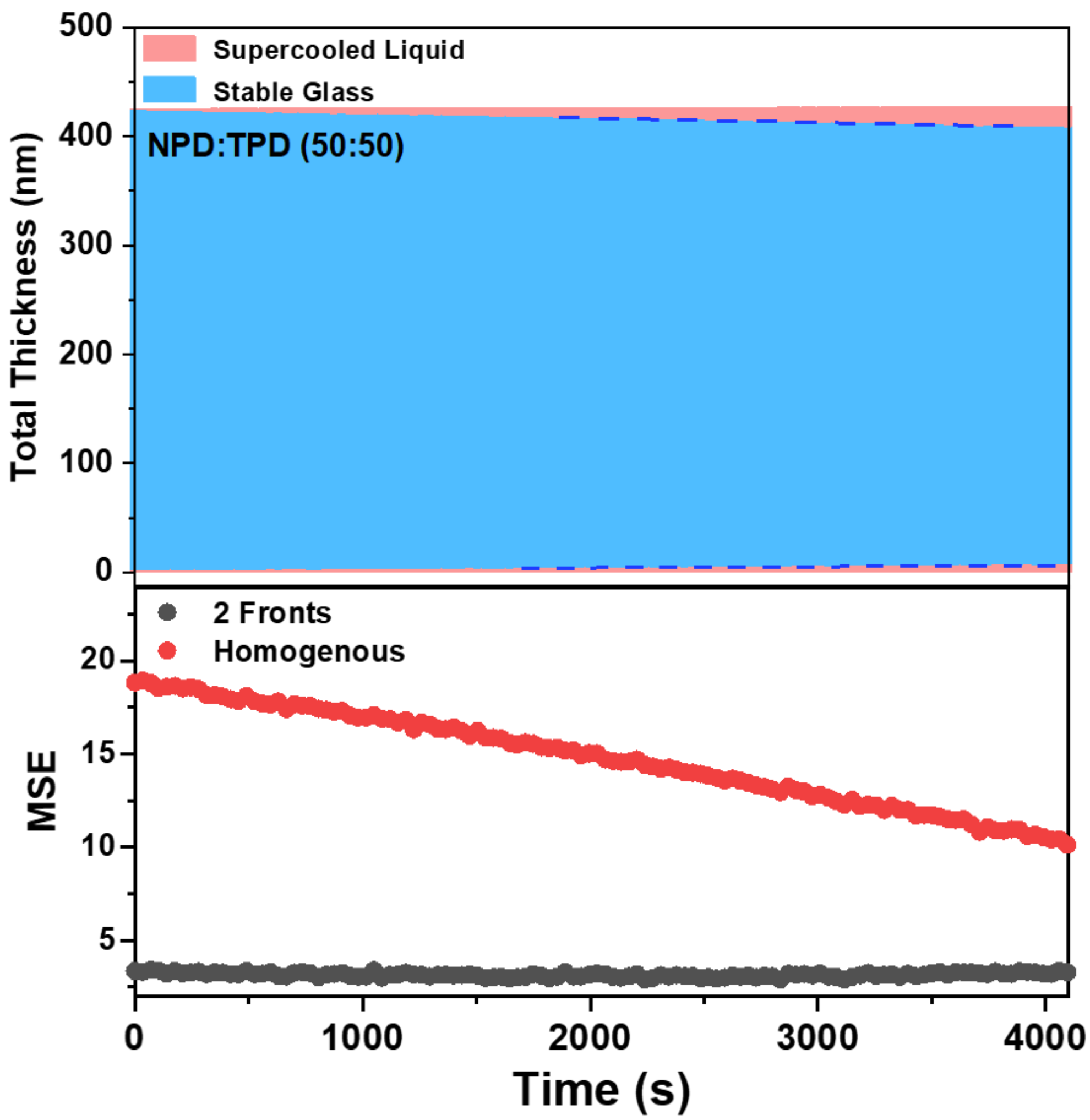

***Figure 5*** *Time evolution of NPD:TPD (50:50) co-deposited at 315 K (0.91 $T_g$) during isothermal annealing at 360 K by ellipsometric measurement (Top) Thickness of stable glass colored in light blue and its transformed supercooled liquid colored in pink. It was fitted with a three-layer model considering surface and substrate-initiated transformation. The dashed lines are guide to eye to show that it is transforming via a front growth at constant velocity. The velocity from surface and substrate initiation is $0.37*10^{-4}$ nm/sec and $0.24*10^{-4}$ nm/sec, respectively. (Bottom) Corresponding MSE at each measurement for two different fitting model for comparison. The homogenous model based on the bulk transformation has higher MSE than 2 fronts model.*

## Supporting Information

Experimental materials and methods; Normalized film thickness for co-deposited NPD:TPD mixtures (Figure S1); Refractive index (n) for liquid cooled NPD:TPD glasses (Figure S2); DSC thermograms of bulk NPD:TPD mixtures (Figure S3); Raw spectroscopic data from ellipsometry and the fitted model over the wavelength (Figure S4)

## Acknowledgment

This work was supported by Department of Energy (DOE), Office of Basic Energy Science, (Grant No. DE-SC0002161). We thank Professor Lian Yu for providing access to the DSC instrument.

# Supporting Information for

# High density two-component glasses of organic semiconductors prepared by physical vapor deposition.

Yejung Lee[1], Shinian Cheng[1], and M. D. Ediger[*,1]

[1]*Department of Chemistry, University of Wisconsin-Madison, Madison, Wisconsin, 53706, USA*

## Materials and Methods

### Sample preparation

TPD (99% purity) and NPD (96% purity) were purchased from Sigma-Aldrich and used as received without further purification. In a vacuum chamber with a base pressure of $10^{-6}$ Torr, the films were prepared by PVD onto a silicon wafer (Virginia Semiconductor, Inc.) with 2 nm of native oxide layer. The silicon substrate was attached to temperature controllable fingers. The total deposition rate was around 0.2 nm/s (0.18 – 0.24 nm/s) and was monitored by a quartz crystal microbalance (QCM). The desired composition of TPD and NPD were attained by separately controlling the deposition rate from two crucibles. For example, in the case of TPD:NPD (10:90) samples, the deposition rate of TPD and NPD was about 0.02 nm/s and 0.18 nm/s, respectively. During deposition, both components were evaporated simultaneously, and a custom-built crucible shield was used to check the deposition rate of each component before and after deposition. In addition, the sample composition was checked by comparing the $T_g$ value of the sample. For each given composition, for separately deposited samples, the $T_g$ values were consistent to ±1 K. This indicates that the composition for the six samples was consistent to within ±0.02.

## Spectroscopic Ellipsometry

A variable angle spectroscopic ellipsometer (VASE) from J.A. Woollam (M-2000U) was utilized with three different procedures: sample mapping at room temperature, temperature ramping and isothermal annealing. For all three procedures, the data was fitted with a uniaxial anisotropic Cauchy model ($n = A+B/\lambda^2 +C/\lambda^4$) within wavelength range of 550 – 1000 nm. The root-mean-squared error (MSE) for the model was determined for all fits. Lower MSE values signify that the model fits well with the data.

To determine the thickness and birefringence ($n_z$-$n_{xy}$), sample mapping was conducted at room temperature at three incidence angles (50°, 60° and 70°). Subsequently, for the temperature ramping experiment, the sample was heated and cooled at 5 K/min for 3 cycles onto a house-built ramping stage equipped with ellipsometer while performing measurements at a single angle of incidence of 70°. These dilatometry experiments determined the onset temperature ($T_{onset}$), glass transition temperature ($T_g$) and fictive temperature ($T_f$). To identify the relative density, the thickness of as-deposited and liquid-cooled glasses at 305 K was compared.

For a few samples, isothermal annealing experiments were performed on as-co-deposited glassy films at 360 K. To increase precision, data was collected every 35 seconds at 3 incident angles: 50°, 60° and 70°. The first and last data sets were taken to be the as-deposited glass and the fully transformed liquid, respectively. Intermediate data sets were fitted with a three-layer model consisting of supercooled liquid, stable glass, and supercooled liquid; the optical constants (n, $n_z$-$n_{xy}$, A, B and C values for Cauchy model) were fixed for the stable glass and the supercooled liquid. In this procedure, only the thicknesses of the layers (3 parameters) are determined by fitting. Based upon this, the position of the two transformation fronts can be plotted in Figure 6. For the comparison, a model utilizing a single homogeneous model was fitted as well.

## Differential Scanning Calorimetry (DSC)

To characterize the bulk mixtures of TPD and NPD, thermal analysis using differential scanning calorimetry (DSC) of TA Q2000 was conducted. For the sample preparation, as-purchased crystalline TPD and NPD at various mass ratio were physically mixed and ground using a mortar and pestle. Subsequently, the crystalline mixtures were sealed into DSC pans. With the rate of 10 K/min and nitrogen gas purging 50 mL/min, the samples were heated to above the melting temperature of both materials to melt completely in the first cycle and then subsequent cycles were used to obtain the glass transition temperature of mixtures.

**S1 Ellipsometric measurements of normalized density for as-deposited NPD:TPD glasses at different compositions and the corresponding liquid-cooled glasses, during heating and cooling at 5 K/min. All the data were normalized at corresponding $T_g$ for visual clarity. (a) 100:0 (pure NPD), (b) 90:10, (c) 50:50 and (d) 0:100 (pure TPD)**

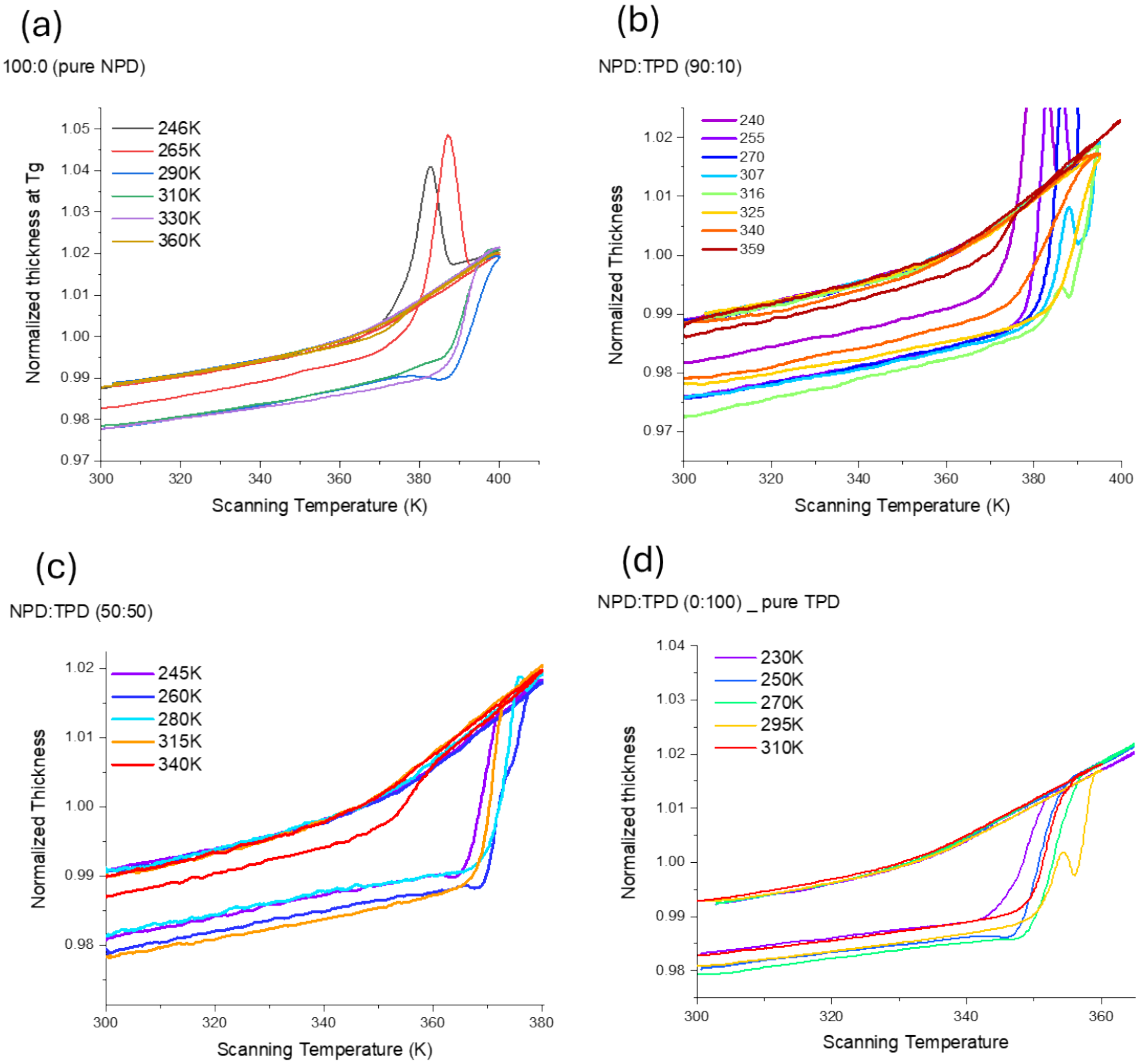

**S2 Refractive index (n) at 650 nm for liquid cooled NPD:TPD glasses during temperature ramping when the data were fitted with isotropic Cauchy model. Color scheme is same as Figure 2. Kink observed in this evolution of refractive index was also for determining glass transition temperature of ordinary glasses, and the values are written with colored arrow.**

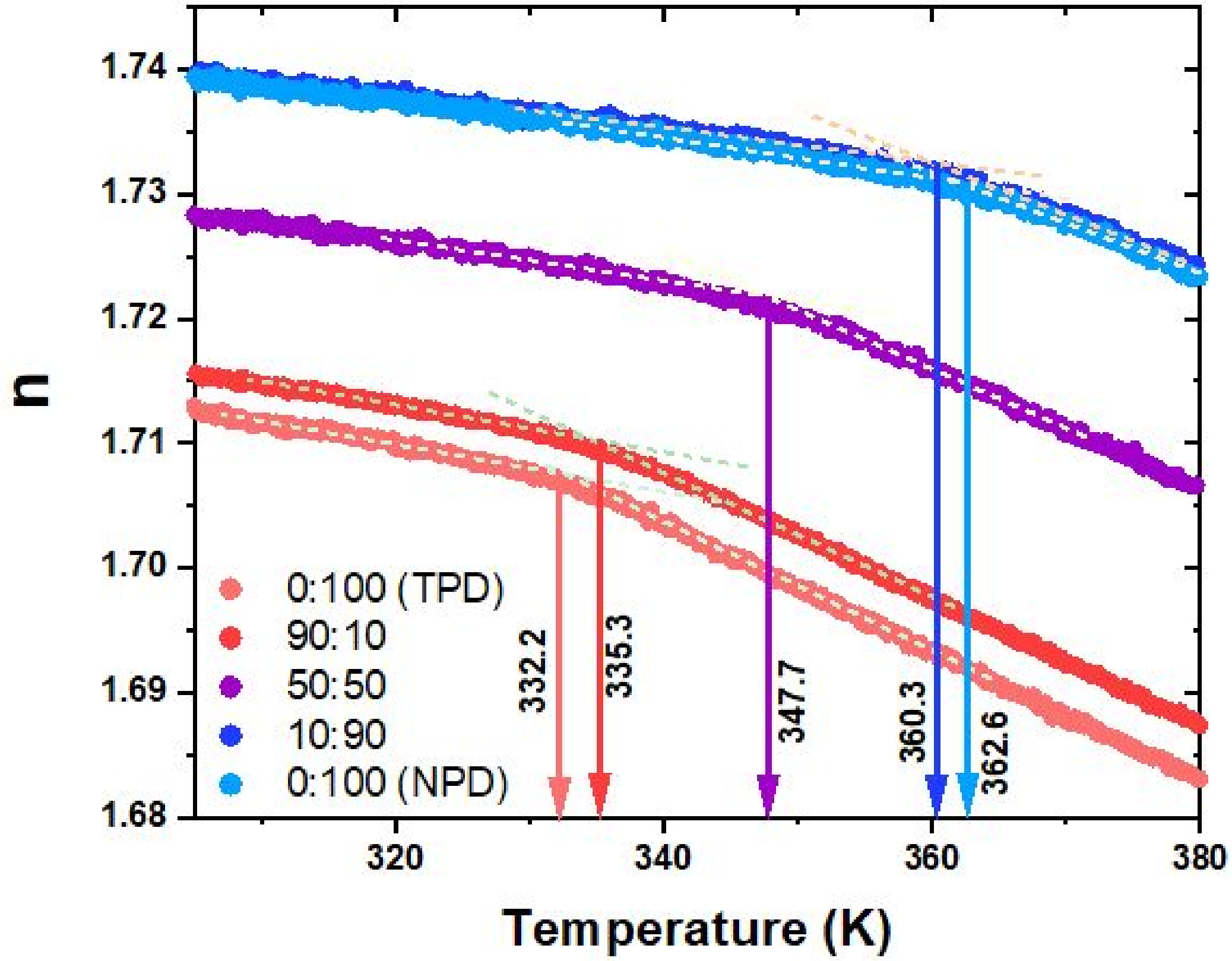

**S3 Differential scanning calorimetry thermograms of bulk mixtures of NPD:TPD (a) melting of the crystalline physical mixture of NPD:TPD. (b) glass transition temperature of the mixture. The mass ratio is indicated in the corresponding color.**

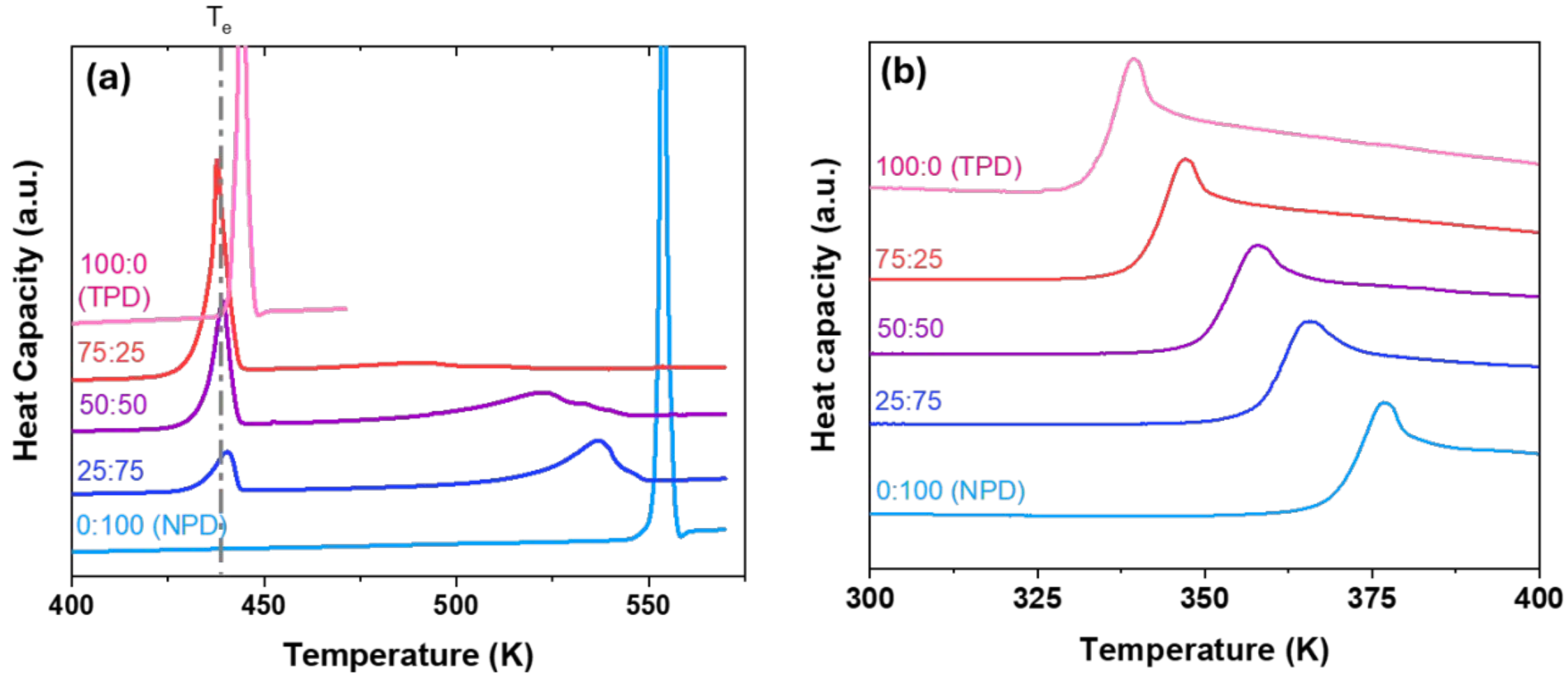

**S4 Spectroscopic data for a 210 nm thick NPD:TPD (50:50) co-deposited glasses at $T_{sub}$ = 283K as a function of wavelength. Raw data of Psi and Delta is shown in red and green and, for comparison, the model is in dotted black. The raw data and the model match well.**

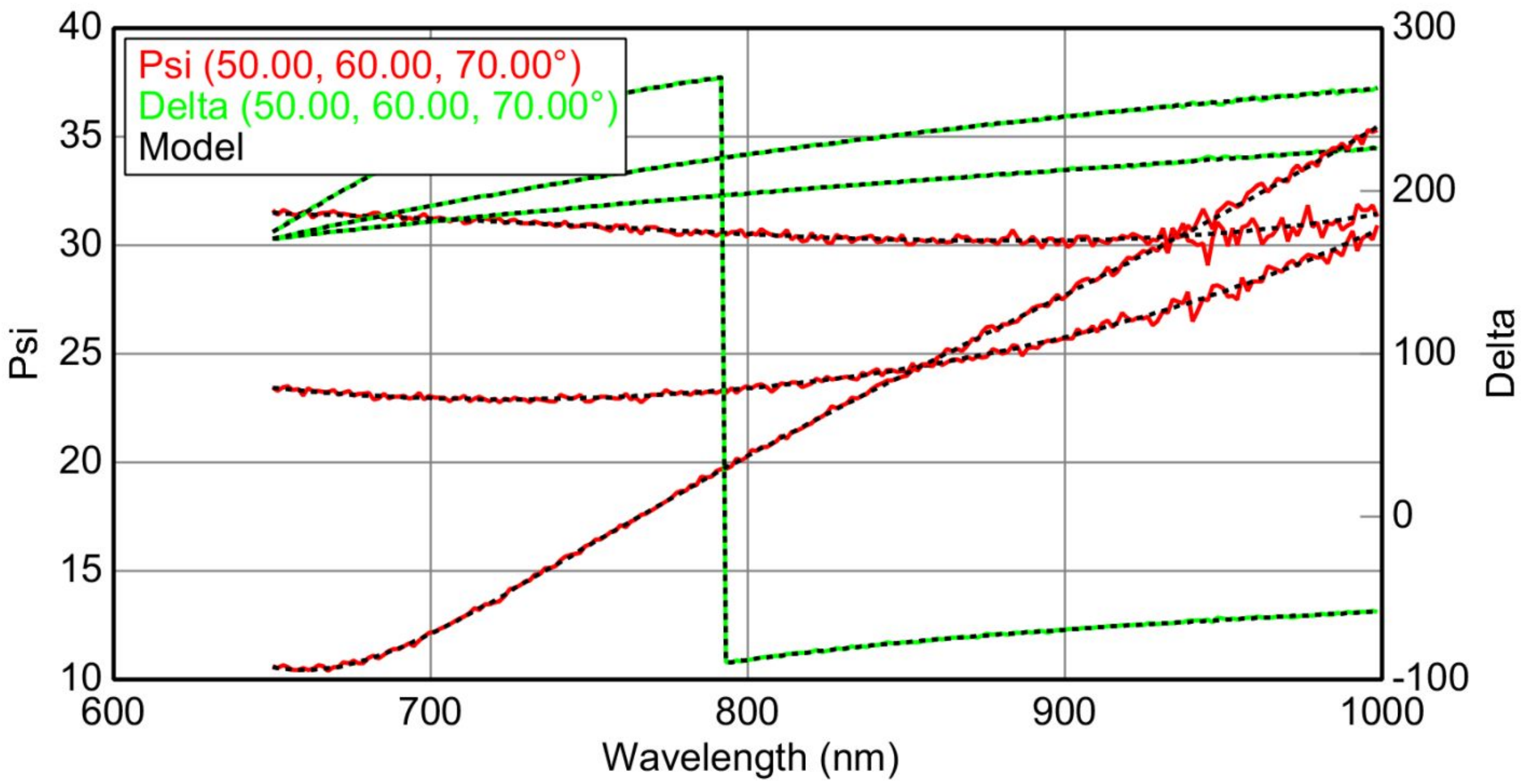

**S5 Refractive indices (n) as a function of wavelength (nm) for two different NPD:TPD (50:50) liquid-cooled glasses as determined by fits to the isotropic Cauchy model. These data were used for Figure S2. The good agreement between these two independent samples indicates a reliable determination of the refractive index.**

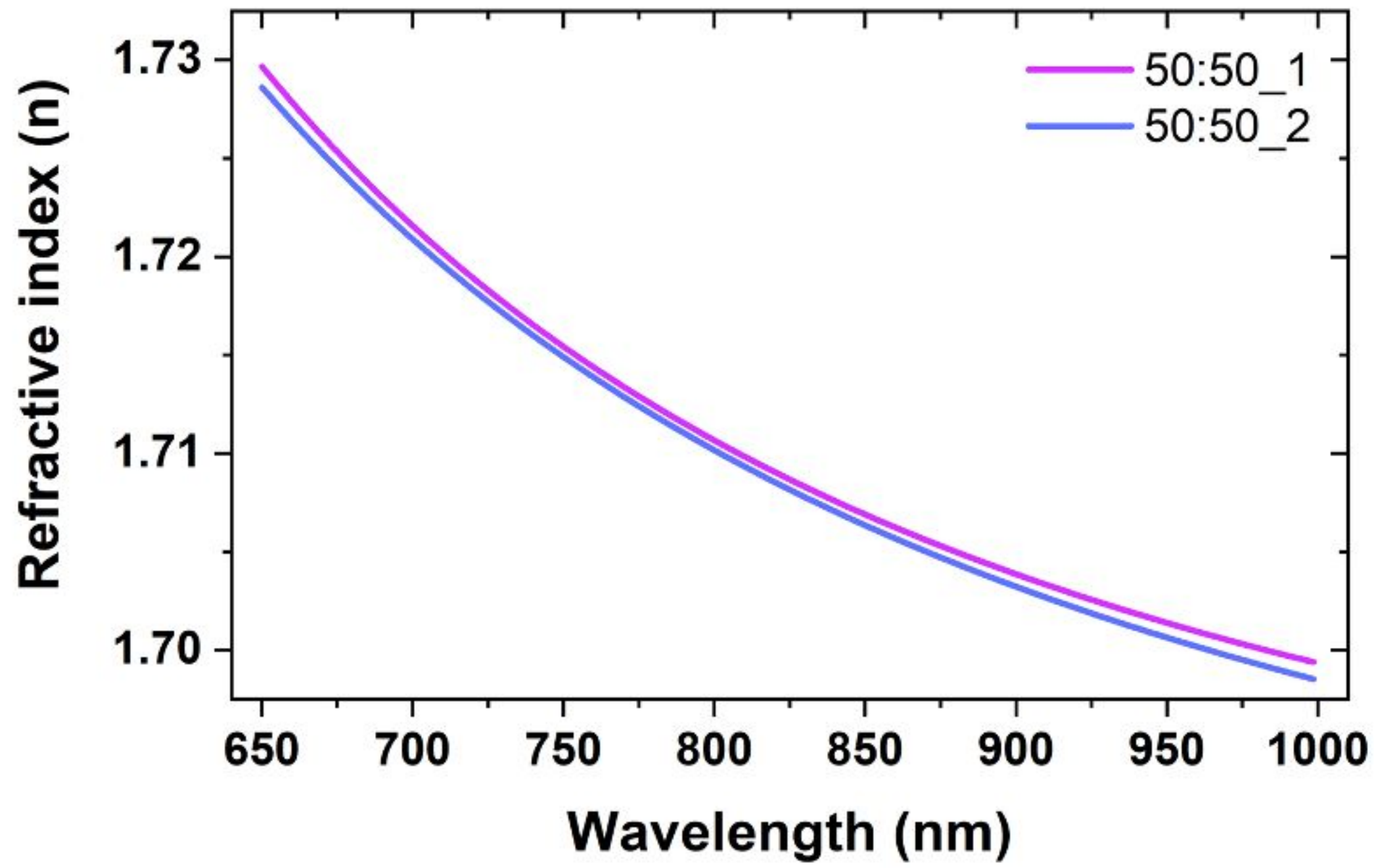